# PERFORMANCE EVALUATION OF DIFFERENT SCHEDULING ALGORITHMS IN WIMAX

Ala'a Z. Al-Howaide[1], Ahmad S. Doulat[2], Yaser M. Khamayseh[3]

Department of Computer Science, Jordan University of Science and Technology, Irbid, Jordan

[1]computergy_alaa@yahoo.com, [2]doulat.ahmad@yahoo.com, [3]yaser@just.edu.jo

## ABSTRACT

*Worldwide Interoperability for Microwave Access (WiMAX) networks were expected to be the main Broadband Wireless Access (BWA) technology that provided several services such as data, voice, and video services including different classes of Quality of Services (QoS), which in turn were defined by IEEE 802.16 standard. Scheduling in WiMAX became one of the most challenging issues, since it was responsible for distributing available resources of the network among all users; this leaded to the demand of constructing and designing high efficient scheduling algorithms in order to improve the network utilization, to increase the network throughput, and to minimize the end-to-end delay.*

*In this study, we presenedt a simulation study to measure the performance of several scheduling algorithms in WiMAX, which were Strict Priority algorithm, Round-Robin (RR), Weighted Round Robin (WRR), Weighted Fair Queuing (WFQ), Self-Clocked Fair (SCF), and Diff-Serv Algorithm.*

## KEYWORDS

*WiMAX,Quality of Services (QoS), Strict Priority (SP), Round Robin (RR), Weighted Round Robin (WRR), Weighted Fair Queuing (WFQ), Self-Clocked Fair (SCF) Queuing, Diff-Serv (DS).*

## 1. INTRODUCTION

WiMAX is a telecommunication protocol that provides fixed and mobile internet access, in which this protocol combines a number of wireless technologies that have emerged from IEEE to face the rapid demand of higher data rate and longer transmission range in wireless access and to enable a high speed connection to the Internet in terms of multimedia service, trade, commerce, education, research and other applications. In other hand, WiMAX technology based on IEEE 802.16 standard which is a Broadband Wireless Access (BWA) that offers mobile broadband connectivity [2].

The rest of this study is organized as follows: section 2 describes the architecture of WiMAX networks, section 3 mentions the QoS classes, several scheduling algorithms are discussed in section 4, the simulation results and analysis are provided in section 5, and section 6 provides some concluding remarks and future work issues.

## 2. WIMAX ARCHITECTURE

WiMAX based on the standard IEEE 802.16, which consist of one Base Station (BS) and one or more Subscriber Stations (SSs), as shown in Figure 1, the BS is responsible for data transmission from SSs through two operational modes: Mesh and Point-to-multipoint (PMP), this transmission can be done through two independent channels: the Downlink Channel (from BS to SS) which is used only by the BS, and the Uplink Channel (from SS to BS) which is shared between all SSs, in Mesh mode, SS can communicate by either the BS or other SSs, in this mechanism the traffic can be routed not only by the BS but also by other SSs in the network, this means that the uplink and downlink channels are defined as traffic in both directions; to and from the BS. In the PMP mode, SSs can only communicate through the BS, which makes the provider capable of monitor the network environment to guarantee the Quality of Service QoS to the customers [1], [11].

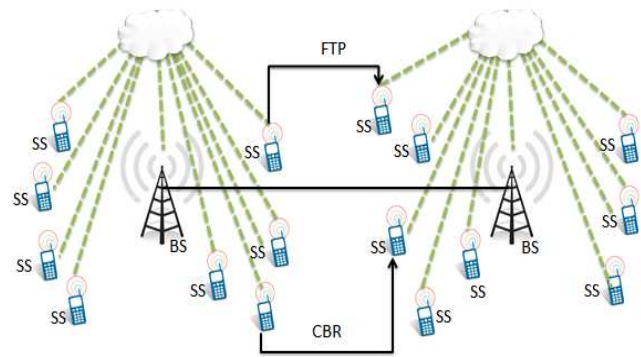

Figure 1. WiMAX Architecture

## 3. QUALITY OF SERVICE (QOS)

QoS parameters are the classes that the BS in a network should support to be able to support a wide variety of applications [10], those parameters include:

- Unsolicited Grant Service (UGS): that supports constant Bit Rate (CBR) such as voice applications.

- Real-Time Polling Service (rtPS): support real-time data streams that contain variable-size data packets, which are issued at periodic intervals such as MPEG video.

- Extended Real-Time Polling Service (ertPS): applicable with variable rate real-time applications that require data rate and delay guarantees like VoIP with silence suppression.

- Non-Real-Time Polling Service (nrtPS): support delay tolerant data streams that contains variable-size data packets, that require a minimum data rate like FTP, and

- Best Effort (BE): support data streams that do not need any QoS guarantees like HTTP.

## 4. WIMAX SCHEDULING ALGORITHMS

Scheduling algorithms are responsible for Distributing resources among all users in the network, and provide them with a higher QoS. Users request different classes of service that may have different requirements (such as bandwidth and delay), so the main goal of any scheduling algorithm is to maximize the network utilization and achieve fairness among all users.

### 4.1. Strict Priority (SP)

In this algorithm packets are represented by the scheduler depending on the QoS class and then they are assigned into different priority queues, these queues are served according to their priority from the highest to the lowest as shown in Figure 2, in which this mechanism may causes some priority QoS classes to be starved.

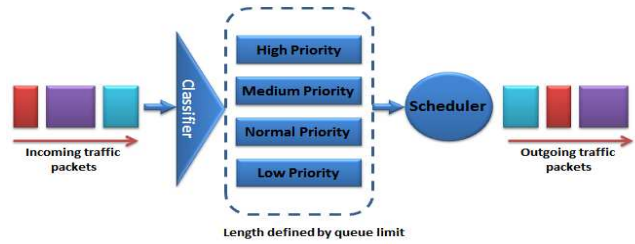

Figure 2. Strict Priority Scheduler

## 4.2. Round Robin (RR)

Figure 3 shows that the procedure of RR scheduler works in rounds by serving the first packet in each priority queue in sequence according to their precedence till all queues are served and then it restarts over to the second packet in each queue.

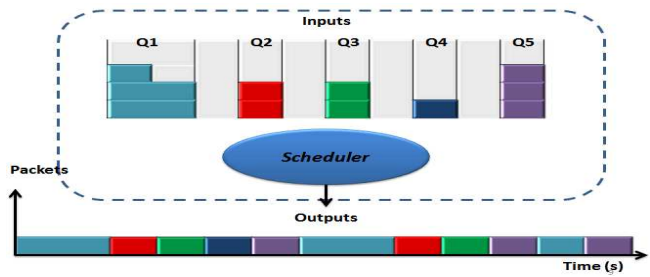

Figure 3. Round Robin Scheduler

## 4.3. Weighted Round Robin (WRR)

In WRR procedure, packets are categorized into different service classes and then assigned to a queue that can be assigned different percentage of bandwidth and served based on Round Robin order as shown in Figure 4. This algorithm address the problem of starvation by guarantees that all service classes have the ability to access at least some configured amount of network bandwidth.

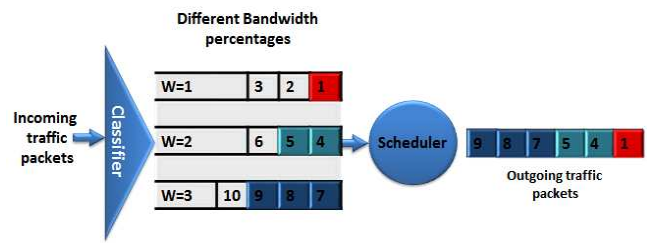

Figure 4. Weighted Round Robin Scheduler

## 4.4. Weighted Fair Queuing (WFQ)

As shown in Figure 5, each flow are assigned different weight to has different bandwidth percentage in a way ensures preventing monopolization of the bandwidth by some flows providing a fair scheduling for different flows supporting variable-length packets by approximating the theoretical approach of the generalized processor sharing (GPS) system that calculates and assigns a finish time to each packet.

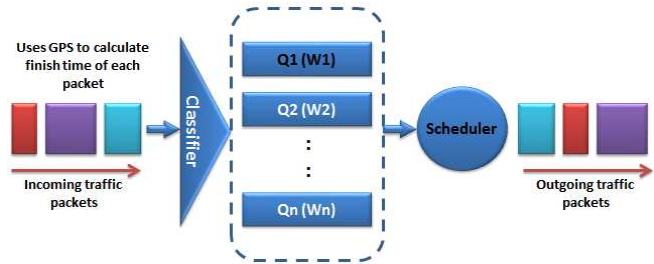

Figure 5. Weighted Fair Queuing Scheduler

## 4.5. Self-Clocked Fair (SCF) Queuing

SCF Scheduler generates virtual time as an index of the work progress; this time is computed internally as the packet comes to the head of the queue. The virtual time determines the order of which packets should be served next, Figure 6 illustrates the work progress of SCF scheduler.

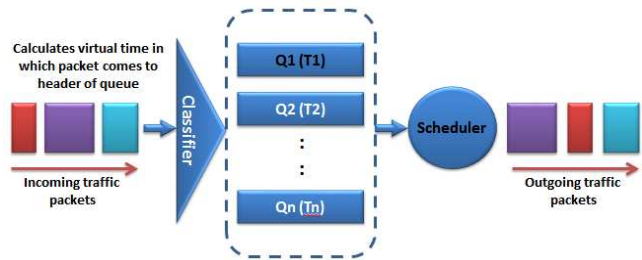

Figure 6. Self-Clocked Fair Scheduler

## 4.6. Diff-Serv (DS) Enabled

Diff-Serv Uses the 6-bit Differentiated Services Code Point (DSCP) field in the header of IP packets that used to classify packets, by replacing the out dated IP precedence with a 3-bit field in the Type of Service byte of the IP header originally used to classify and prioritize types of traffic.

## 5. LITERATURE REVIEW: SCHEDULING ALGORITHMS EVALUATION

Several researchers analysed and evaluated different scheduling algorithms:

In [3], Mohammed Sabri Arhaif evaluated the implementation of various types of scheduling algorithms in WiMAX network, such as Diffserv-Enabled (Diffserv), Round Robin (RR), Self-Clocked-Fair (SCF), Strict-Priority (SP), Weighted-Fair Queuing (WFQ) and Weighted-Round Robin (WRR). In this study QualNet 5.0 simulator evaluation version are used to evaluate these algorithms and to determine the most efficient one among them.

The system parameters in the simulation consists of a single BS and a number of Mobile Stations (MSs), varies from 10 to 50 MSs, the BS radius range is 1000 meters, MS radius range is 500 meters, the frequency band is 2.4 GHz, the channel bandwidth is 20 MHz, frame duration of 20 ms, the Fast Fourier Transform (FFT) size is 2048, the BS transmission power is 20/5 P_t dBm/height (m), the MS transmission power is 15/1.5 P_t dBm/height (m), and a simulation time of 30 seconds, the QoS parameters that the simulation covers are BE, nrtPS, rtPS, ertPS, and UGS.

Six experiments with different parameters were carried out; the results showed that the SP, WRR and WFQ are more efficient in terms of end-to-end time delay, the behaviour of algorithms were widely different when the number of MS was small (10 MS), RR dominated other algorithms when the number of MSs became more than 50, SCF performed better than Diffserv, WRR, SP, and WFQ when the number of MS became more than 40, RR algorithm achieved the highest value of throughput when the number of MS was more than 30, WFQ showed the best performance as the average end-to-end time delay had the lowest reading. Another observation that the RR algorithm was the most efficient in terms of overall throughput 125Kbps, SP and WRR had the shortest amount of end-to-end delay time for all classes of QoS, RR algorithm achieved the best percentage of fairness index. And as a conclusion of this evaluation, the best scheduling algorithms were: WF, in terms of the amount of end-to-end delay. RR algorithm was the best in terms of packet latency (Jitter). Finally WRR outperformed the rest scheduling algorithms by producing the highest rate of throughput of data packet in the network.

In [4], Ashish Jain and Anil K. Verma descried three scheduling algorithms which were: Proportionate Fair (PF) Scheduling [5], Cross-Layer Scheduling Algorithm [6] and TCP-Aware Uplink Scheduling Algorithm for IEEE 802.16 [7]. And it was proposed to provide a comparative study of these algorithms to define the pros and cons for each technique. First for PF algorithm which had the advantage of fairness in scheduling priority based, and a simple implementation multi-user diversity gain, but in this algorithm no QoS parameters were guaranteed. In other hand, Cross-Layer algorithm guaranteed the QoS parameters, and the channel quality was considered in the scheduling, but it had a complex implementation and all slots per frame were allocated to the highest priority connection. And finally the TCP-aware uplink algorithm which was efficient in utilizing the resources among BE connections, but this was not enough to treat with only one class of QoS, and it has a complex implementation.

In [8], Ahmed Rashwan, Hesham ElBadawy, and Hazem Ali performed a detailed simulation study, in addition to analysing and evaluating the performance of some scheduling algorithms, which were WFQ, Round Robin, WRR and Strict-Priority. The simulation experiments were performed using QualNet version 4.5 evaluation version.

The system parameters in their simulation consisted of five MHz bandwidth with 512, the Fast Fourier Transform (FTT) size was configured to simulate bandwidth congestion in order to study the effect of the heavy traffic on each QoS class with different scheduling algorithms, a transmission parameter with TX-power of 15 dBm were used, channel bandwidth of 5 MHz, FFT size of 512, cycle prefix of 8, frame duration of 20ms and TDD duplex mode, and the parameters for the BS were: OMNI antenna type, 15 dB antenna gain, and 25ms antenna height, eight queues were configured to avoid queuing packets of different service types into one queue. And the precedence for each class of QoS is: BE of 0, nrtPS of 2, rtPS of 3, ertPS of 4 and UGS of 7.

The simulation results showed that the UGS, ertPS and rtps traffic had the largest throughput value. However the BE and nrtPS traffic almost had no traffic because the Strict-Priority scheduler caused bandwidth to be starved for low priority traffic types, the higher priority traffic had a higher throughput and the lowest priority traffic had low throughput, meanwhile WRR distributed the bandwidth according to the assigned weights to all traffic types, WFQ and WRR were very similar despite that they were different in distributing the bandwidth among the traffic types, Strict-Priority scheduler produced the highest UGS, rtPS traffic against the speed since it serves the highest priority traffic queues, RR was fair algorithm but this make it degrade the UGS, rtPS throughput to approximately half of the Strict-Priority, at the same time it increased the BE, nrtPS to be double more, RR scheduler had equal average end-to-end delay for all traffic types except for the BE it had a higher value. RR scheduler had also better performance for low QoS classes on the expense of the high QoS classes. Both WFQ and WRR can control the performance of each class by assigning different weight to each queue.

In [9], Jani Lakkakorpi, Alexander Sayenko and Jani Moilanen presented a detailed performance comparison of some scheduling algorithms such as Deficit Round-Robin, Proportional Fair and Weighted Deficit Round-Robin, taking into account in their comparison the radio channel conditions and the throughput improvement was considerable. The simulation experiments were obtained on a modified version of ns-2 simulator [13], conducting several numbers of simulations for each case of the study to assure 95% confidence interval and a simulation time of 200 seconds. One-way core network delay was set to 31 ms.

The traffic mix was simulated, having 5 VoIP connections, 5 video streaming connections (DL only); 10, 14, 18, 22, 26 or 30 web browsing connections and 5, 7, 9, 11, 13 or 15 file downloading connections per BS. All user traffic was given BE treatment except for VoIP traffic that was given rtPS treatment.

The network parameters used in the simulations: PHY is OFDMa, and the duplexing mode was TDD, a frame length of 5 ms, the bandwidth used was 10 MHz, FTT size of 1024, cyclic prefix length was 1/8, the Transmit-receive Transition Gap was 296 PS, the Receive-transmit Transition Gap was 168 PS, the DL/UL permutation zone was FUSC/PUSC with ratio 35/12, the DL-MAP/UL-MAP fixed overhead was 13 bytes/ 8 bytes, and one opportunity as a number of ranging, ranging back-off start/end was 0/15, three opportunities as a number of requests, request back-off start/end was 3/15, the CDMA codes for ranging and BW requests of 64/192, the maximum size for MAC PDU was 100 bytes, the fragmentation and packing were taken into account, all connections but VoIP with ARQ (Automatic Repeat reQuest) [14], and all the connections were with ARQ feedback types, the ARQ block size of 16 bytes, and the window size of 1024, and there was no ARQ block arrangement.

The simulation resulted in the fact that both PF and WDRR algorithms performed better than DRR in terms of MAC throughput and TCP good-put, the WDRR had a good performance in time this scheme was easier to implement and less computationally complex than PF, meanwhile the PF scheduler can leave a connection without any resources for a long time period that if it was large enough make a problem if ARQ times were set to expire in short time, in other hand, the differences of round-trip time RTT may lead to retransmissions of TCP, that make it possible to the TCP good-put to be degraded, WDRR scheduler performed better than PF when the traffic load was small, since the PF algorithm needs to have enough connections to achieve throughput gain, and by increasing the number of connections the PF algorithm picked the connections with a good MCS, however when the time reserved for connections without resources was large in the PF scheduler to had a better TCP good-put cause increasing in delay, finally the results showed that when the Active Queue Management AQM at the BS was used, it causes the queuing delay to be reduced without affecting the good-put.

## 6. SIMULATION

All the experiments are developed and run using Qualnet V 5.0 simulator using IEEE MAC802.16 protocol. A QualNet is a commercial network simulation tool implemented in C++ that simulates wireless and wired packet mode communication networks. QualNet used in the simulation of MANET, WiMAX networks, satellite networks, wireless sensor networks, etc. It has a graphical user interface and a sets of library function used for network communication [12].

### 6.1. Simulation Model

The purpose of these simulation experiments is to study the impact of Queue size and number of queues within the BS of a WiMAX network on the scheduling algorithms proposed by IEEE for 802.16 protocols.

The WiMAX network is simulated by consisting one BS and 40 SS which are distributed around the BS as shown in Figure 7 with different distances from the BS, while the distance does not affect the scheduling algorithms performance [1].

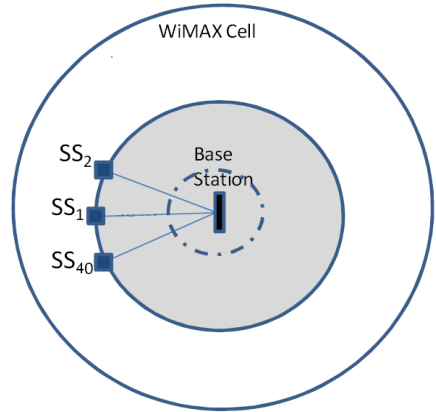

Figure 7. Simulation Environment

The simulation parameters that we are used in our experiments are presented in Table 1.

Table 1. Simulation Parameters

| Parameter | Value |
|---|---|
| Channel bandwidth (MHz) | 2.4 |
| Number of SS | 40 |
| Number of BS | 1 |
| BS transmit power (dBm)/ Height (m) | 20/5 |
| SS transmit power (dBm)/Height (m) | 15/1.5 |
| Services types (QoS) | BE, nrtPS, rtPS, ertPS, UGS |
| Simulation time (s) | 30 |

To simulate the different QoS types a mapping with different precedence values are used as shown in Table 2, which shows this mapping.

Table 2. MAC Layer Service Flow Mapping MAC Layer Services Precedence

| QoS | Precedence Value |
|---|---|
| BE | 0 |
| nrtPS | 1 , 2, 6 |
| rtPS | 3 |
| ertPS | 4 |
| UGS | 5, 7 |

## 6.1. Simulation Scenarios

Two main simulation scenarios are used in an effort to evaluate the effect of BS queue size and number with 6 different scheduling algorithms. The first scenario is carried on by changing the BS output queue size, where three different values are used (128000, 1280000, 12800000 byte). These values are selected in order to test the scheduling algorithms performance when using an output queue size can handle less number of packets received, exact number of packets received, more than received packets, respectively. The second scenario is carried on to test the impact of BS output queue number on the scheduling algorithms performance. Three different values are used (6, 8, 10 output queues).

## 6.2. Simulation Results

The results of the first scenario are presented in Figures 2 to 7, while Figures 8 and 9 show the results of the second scenario.

For the first scenario, one of the most important measures to be found is the server throughput. Figure 8 shows that the performance of all scheduling algorithms is not affected by changing the BS output queue size. DS and WF show the highest server throughput, while SP and SCF show the lowest throuout.

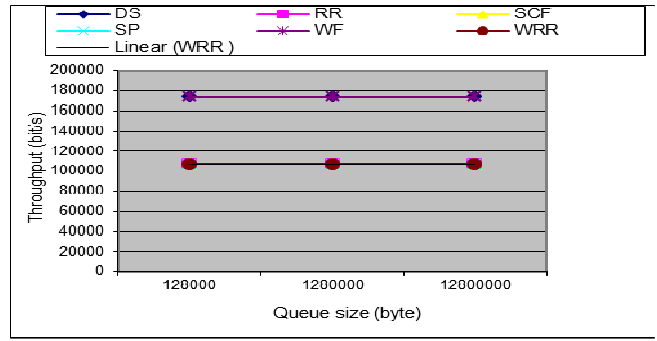

Figure 8. Server Throughput (bit/s) VS Queue size

Another important measure to find is the end-to-end delay, which is showed by Figure 9. The end-to-end delay measure for all scheduling algorithms is not changed for all the three different BS output queue values. The highest delay is 2.96802 sec consumed by DS and WF, where the lowest delay is consumed by RR and WRR with a value of 2.28231 sec.

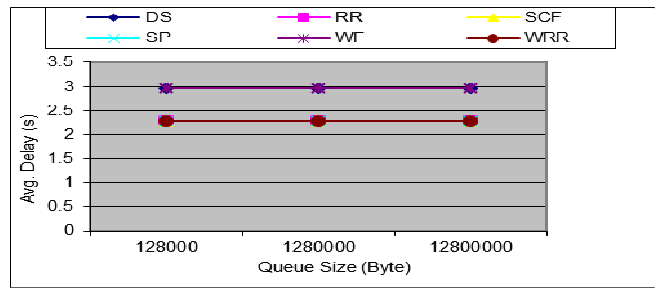

Figure 9. Average End-To-End Delay (s) VS Queue size

Figure 10 shows the peek BS queue size. The average peek queue is increased as the output queue increased. The worst scheduling algorithms at queue size 128 KB are DS, SP, and WF with value of 29712.08 byte, while the best with value 29329.58 byte are RR, WRR, and SP. At queue size 120 KB, still DS, SP, and WF are the worst. While at queue size 12800 KB, the worst is SP with value of 1822332 byte. And the best is DS and WF with value 1800252. The peek queue size of SCF is 1821950 byte, while for RR and WRR is 1821890 byte.

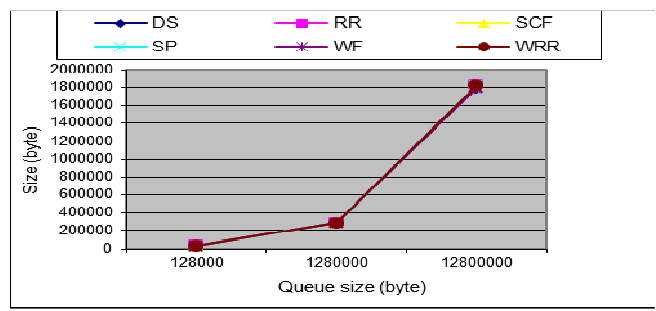

Figure 10. Peak Queue Size (byte) VS Queue size

Average output queue size results are presented in Figure 11, where the highest queue size for all of the three difference queue size values is SP with values 21200.66, 190307.5, and 658712.4, in order. While the lowest average queue size for all the three difference queue size values are DS and WF with values of 20952.43, 187821.6, 642978.3 bytes, in order.

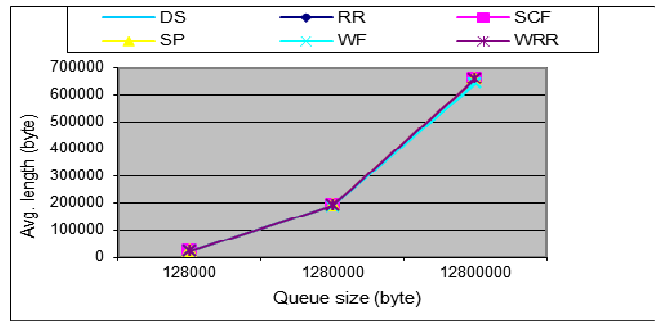

Figure 11. Average Output Queue Length VS Queue size

Figure 12 shows the average time in the queue, which is not affected for all the three different queue size values. The longest average time in the queue is measured at SP with value of 1.508258 sec, while the shortest is at RR, WRR, and SCF with value of 1.504912 sec.

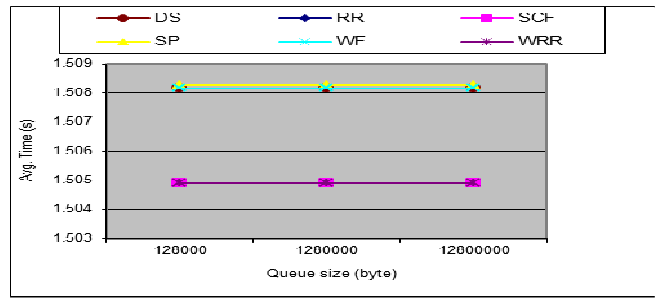

Figure 12: Average Time in Queue (s) VS Queue size

From Figure 10, Figure 11, and Figure 12 we can notice that all scheduling algorithms have nearly a stable output queue growth. In addition, one can find that in average the best scheduling algorithm in output queue management are RR and WRR. They show the highest stability of output queue growth and service time over the three different queue size values.

Figure 13 presents the total dropped packets, which decreased as the output queue size increase. The worst is SP for all the three output queue size values with total dropped packets values 3403.848, 2929.848, and 84.18156 packets, in order. The least total dropped packets is shown by RR and WRR with values of 3353.585, 2879.585, 34.02978 packets, respectively.

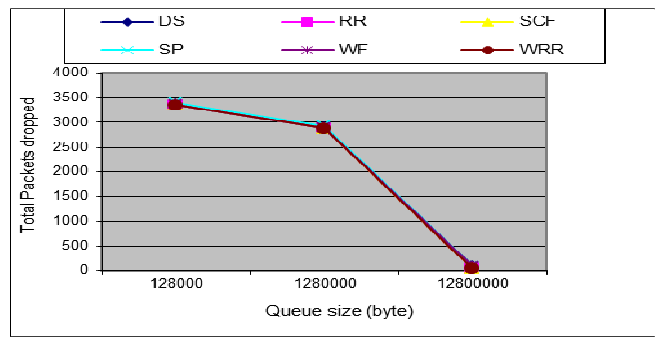

Figure 13. Total Packets Dropped VS Queue size

The average server throughput in the second scenario is shown in Figure 14, the average server throughput increases as the number of output queues increases. The least throughput is produced by DS and WF for all the three different number of queues with the value 173903 bit/sec. The highest throughput is produced by SCF and SP for all the three different number of queues with the value 106933 bit/sec. WRR showed better throughput than RR at 10 output queues with value 106933 bit/sec.

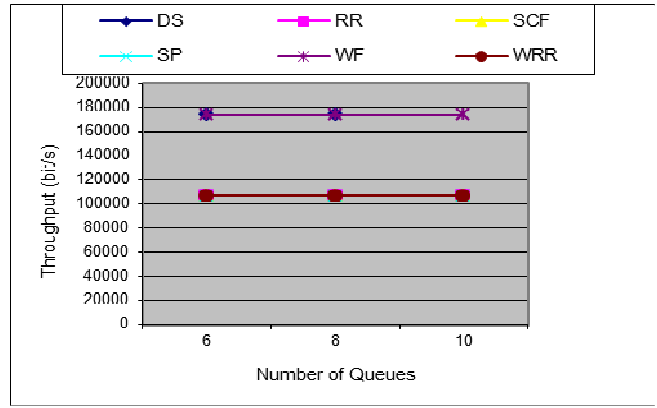

Figure 14. Server Throughput (bit/s) VS Number of Queues

The average time in the queue increases as the number of queues increase as presented in Figure 15. SP is the worst for all the three different number of queues with values 1.007078, 1.007078, 1.007078 sec, in order. RR, WRR, and SCF are the best for all the three different number of queues with values 1.00421, 1.289924, 1.578308 sec, in order.

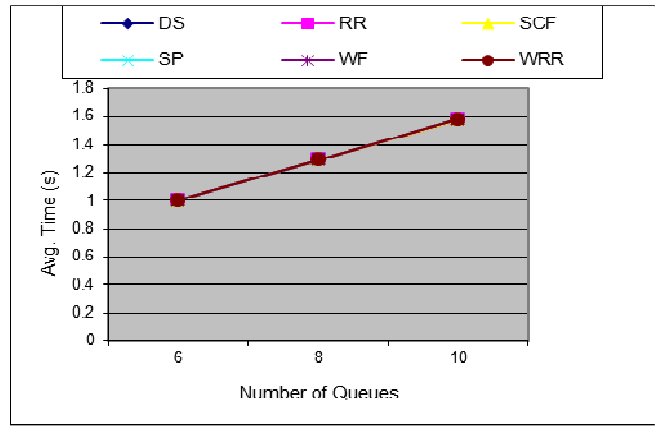

Figure 15. Average Time in Queue (s) VS Number of Queues

## 7. CONCLUSION AND FUTURE WORK

The performance of Strict Priority, Round Robin, Weighted Round Robin, Weighted Fair Queuing, Self-Clocked Fair Queuing, and Diff-Serv scheduling algorithms is measured mainly in terms of size and number of BS output queues within WiMAX network. The five QoS classes are included in the simulation. The results showed that the output queue size and number do not affect the server throughput and end-to-end delay for a specific scheduling algorithm. In addition, the best scheduling algorithms with queue management and resource utilization are RR, WRR, SCF, WFQ, and DS in order.

Developing a new scheduling algorithm which support different queue sizes for different QoS is really necessary and it will be considered as a future work in addition to, performing further tests on the impact of queue priorities (all different, same, or clusters).

## ACKNOWLEDGMENT

We would like to thank professor Wail Mardini for his support to make this work done and his most helpful notes.

## Authors

Yaser Khamayseh was born in Irbid/ Jordan in 1977. He received his Bachelor degree in computer science from Yarmouk University, Irbid, Jordan, in 1998. He finished his master's in computer science at the University of New Brunswick, Canada in 2001. And, he finished his PhD in computer science at University of Alberta, Canada in 2007.

He is an assistant professor of computer science at Jordan University of Science and Technology since 2007. He has More than 12 years of experience in research and teaching in the field of data communication and computer networks. His research interests include simulation and modeling, wireless networks, performance evaluation, evolutionary computation, and image processing.

Dr. Yaser Khamayseh is member of IEEE and has received several awards. He is a member of technical programs of several journals and conferences. He served as a reviewer for several conferences and Journals.

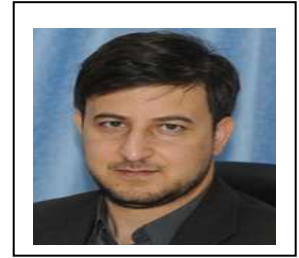

Ahmad S. Doulat received his B.Sc degree in Computer Science from Yarmouk University, Irbid, Jordan in 2007. He is an M.Sc student in computer science in Jordan University of Science and Technology, Irbid, Jordan. His research interests include WiMAX Networks, Wireless Sensor Networks and Cloud Computing.

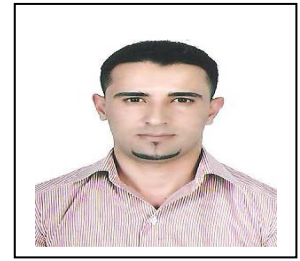

Ala'a Z. Al-Howaide received his B.Sc Degree in Computer Information Systems with honour from Yarmouk University, Irbid, Jordan in 2008. He was the first on his colleges. He worked as an Oracle programmer and web service developer, until he joined the M.Sc in computer science program in Jordan University of Science and Technology, Irbid, Jordan in 2009. His research interests include WiMAX Networks, Intrusion Detection Systems (IDS), Data Mining, and Speech Recognition.

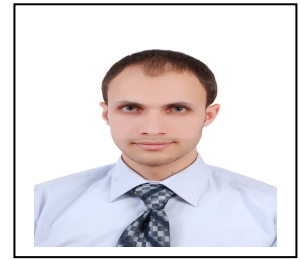